\documentstyle[12pt,twoside]{article}
{\catcode `\@=11 \global\let\AddToReset=\@addtoreset}
\AddToReset{equation}{section}

\def\greaterthansquiggle{\raise.3ex\hbox{$>$\kern-.75em\lower1ex\hbox{$\sim$}}}
\def\lessthansquiggle{\raise.3ex\hbox{$<$\kern-.75em\lower1ex\hbox{$\sim$}}}
\newcommand{\beq}{\begin{equation}}
\newcommand{\eeq}{\end{equation}}
\newcommand{\beqa}{\begin{eqnarray}}
\newcommand{\eeqa}{\end{eqnarray}}
\newcommand{\beqan}{\begin{eqnarray*}}
\newcommand{\eeqan}{\end{eqnarray*}}
\newcommand{\ba}{\begin{array}}
\newcommand{\ea}{\end{array}}
\newcommand{\no}{\nonumber}

\newcommand{\ra}{\rightarrow}

\newcommand{\ve}{\varepsilon}

\newcommand{\B}{{\cal B}}

\newcommand{\Ha}{{\cal H}}

\newcommand{\cO}{{\cal O}}

\newcommand{\R}{{\cal R}}

\newcommand{\st}{\stackrel}

\def\nz{\ifmmode {I\hskip -3pt N} \else {\hbox {$I\hskip -3pt N$}}\fi}
\def\zz{\ifmmode {Z\hskip -4.8pt Z} \else
       {\hbox {$Z\hskip -4.8pt Z$}}\fi}
\def\qz{\ifmmode {Q\hskip -5.0pt\vrule height6.0pt depth 0pt
       \hskip 6pt} \else {\hbox
       {$Q\hskip -5.0pt\vrule height6.0pt depth 0pt\hskip 6pt$}}\fi}
\def\rz{\ifmmode {I\hskip -3pt R} \else {\hbox {$I\hskip -3pt R$}}\fi}
\def\cz{\ifmmode {C\hskip -4.8pt\vrule height5.8pt\hskip 6.3pt} \else
       {\hbox {$C\hskip -4.8pt\vrule height5.8pt\hskip 6.3pt$}}\fi}

\def\au{{\setbox0=\hbox{\lower1.36775ex%
\hbox{''}\kern-.05em}\dp0=.36775ex\hskip0pt\box0}}
\def\ao{{}\kern-.10em\hbox{``}}

\normalsize
\begin{document}
\bibliographystyle{plain}
\begin{titlepage}
\begin{flushright}
UWThPh-1996-37\\
\today
\end{flushright}
\vspace{2cm}
\begin{center}
{\Large \bf  TT-tensors and conformally flat structures \\
on 3-manifolds}\\[50pt]
R. Beig*  \\
Institut f\"ur Theoretische Physik \\ Universit\"at Wien\\
Boltzmanngasse 5, A-1090 Wien 
\vfill
{\bf Abstract} \\
\end{center}

We study transverse-tracefree (TT)-tensors on conformally flat 3-manifolds
$(M,g)$. The Cotton-York
tensor linearized at $g$ maps every symmetric tracefree tensor into one which 
is TT. The question as to whether this is the general solution to the
TT-condition is viewed as a cohomological problem within an elliptic
complex first found by Gasqui and Goldschmidt and reviewed in the present
paper. The question is answered affirmatively when $M$ is simply
connected and has vanishing 2nd de Rham cohomology.

\begin{center}
Talk given at the Workshop ``Mathematical aspects of theories of
gravitation'', \\
Stefan Banach International Mathematical Center, Warsaw, \\ 26 February --
30 March 1996
\end{center}
\vfill \noindent
*) Supported by Fonds zur F\"orderung der wissenschaftlichen Forschung,
Project P9376-MAT.
\end{titlepage}

\section{Introduction}
In the context of the initial-value problem for the Einstein equations
(see [5]) one is often interested in the following problem. Let
$(M,g)$ be a connected, smooth, 3-dimensional, orientable manifold and let
$t_{ab}$ be an element of $S_0^2(M,g)$, that is to say a 2-covariant,
symmetric tensor field which is tracefree with respect to $g_{ab}$,
i.e. $t_{ab} = t_{(ab)}$ and $t_{ab} g^{ab} = 0$, where $g^{ab}$ is the 
inverse of $g_{ab}$. We want to solve the equation
\beq
(\delta t)_a := 2 g^{bc} D_c t_{ab} = 0,
\eeq
where $D_a$ is the Levi-Civita connection associated with $g_{ab}$. 
Elements of $S_0^2(M,g)$ satisfying (1.1) are also called TT-tensors.
The equation (1.1) is an underdetermined elliptic system.
This means that the principal symbol of $\delta$, namely the linear map
$$
\bar\delta(k;x): \tau \in S_0^2({\bf R}^3,g_x) \ra \omega \in \Lambda^1
({\bf R}^3), \qquad k \in \Lambda^1({\bf R}^3), \qquad k \neq 0,
$$
defined by
\beq
\omega_a = g^{bc}(x) k_c \tau_{ab},
\eeq
is surjective.\footnote{That this is the case follows by setting
$$
\tau_{ab} = \frac{2}{k^2} k_{(a} \omega_{b)} - \frac{1}{2k^2} g_{ab}
(k,\omega) - \frac{1}{k^4} k_a k_b(k,\omega),
$$
where $k^2 := g^{ab}(x) k_a k_b$ and $(k,\omega) := g^{ab}(x)\omega_a
k_b$.} There is a general method (see the Appendix of [2]) to solve such
a system, as follows: Define the operator $L$
$$
L : \Lambda^1(M) \ra S_0^2(M,g)
$$
by
\beq
(LW)_{ab} = D_a W_b + D_b W_a - \frac{2}{3} g_{ab} D^c W_c.
\eeq
Clearly $-L = \delta^*$, i.e. $L$ is minus the formal adjoint of $\delta$
under the inner product given by the Riemannian volume element of $g$.
The kernel of $L$ is the finite-dimensional space of covector fields
$W_a$, so that $W^a = g^{ab} W_b$ is a conformal Killing vector field
on $(M,g)$. Furthermore there is the decomposition
\beq
S_0^2(M,g) = L(\Lambda^1(M)) \oplus \ker \delta.
\eeq
Starting with an element $Q_{ab} \in S_0^2(M,g)$, its component 
$t_{ab}$ in $\ker \delta$ can formally be written as
\beq
t = [{\bf 1} - L(\delta \circ L)^{-1} \delta]Q.
\eeq
Since $\ker (\delta \circ L) = \ker L$ and $\delta Q$ is orthogonal to
$\ker L$, the right-hand side of Equ. (1.5) is well defined. The
relations given by (1.3) and (1.4) furnish what is called the York
decomposition (after [19], see also [6]) in the G.R. literature. 
This decomposition
is closely related to the study of the action of conformal diffeomorphisms
on the space of Riemannian metrics on $M$ [9].
In the present work we seek a refinement of this decomposition in a
sense which is best explained by the example of the de Rham--Hodge theory.
Consider, thus, instead of (1.1), the equation
\beq
\mbox{div } \omega = D^a \omega_a = 0.
\eeq
Again, this is an underdetermined elliptic system, and we have the
orthogonal decomposition
\beq
\Lambda^1(M) = \mbox{grad }(C^\infty(M)) \oplus \ker \mbox{div},
\eeq
where grad is minus the formal adjoint of div, namely the differential
acting on functions. Sometimes the relation (1.7) is called Helmholtz
decomposition in the physics literature. The splitting given by (1.7)
can be refined by noticing that there is a large class of explicit
solutions to (1.7) namely all elements $\omega \in \Lambda^1(M)$ of
the form $\omega = \mbox{rot }\mu$, where rot:~$\Lambda^1(M) \ra 
\Lambda^1(M)$ is defined by 
\beq
\omega_a = \ve_a{}^{bc} D_b \mu_c.
\eeq
Every element in grad~$(C^\infty(M))$, in turn, is in the kernel of
rot. Then consider the sequence of spaces and linear maps
\beq
0 \ra C^\infty(M) \st{\rm grad}{\longrightarrow} \Lambda^1(M)
\st{\rm rot}{\longrightarrow} \Lambda^1(M) \st{\rm div}{\longrightarrow}
C^\infty(M) \ra 0.
\eeq
This is a {\bf complex}, i.e. every element in each of these space which
is in the image of the map to the left, is also in the kernel of the
map to the right. It is also an {\bf elliptic complex}, i.e. the
corresponding complex of symbols is {\bf exact}: every element in the
kernel of a symbol map to its right is in the image of the symbol map
to its left. For grad and div this just amounts to the statement that
div is underdetermined elliptic, and, equivalently, that grad is
overdetermined elliptic (the associated symbol map is injective). 
Now define the Hodge Laplacian $\Delta_H$
$$
\Delta_H : \Lambda^1(M) \ra \Lambda^1(M)
$$
by
\beq
\Delta_H = \mbox{(rot)}^2 - \mbox{grad div}.
\eeq
This has the following properties: It is formally self-adjoint and
elliptic (i.e. its symbol is injective and surjective).
Thus (see Warner [18]) $\ker \Delta_H$ is finite-dimensional and
\beqa
\Lambda^1(M) &=& \Delta_H(\Lambda^1(M)) \oplus \ker \Delta_H \no \\
&=& \mbox{grad }(C^\infty(M)) \oplus \mbox{rot }(\Lambda^1(M))
\oplus \ker \Delta_H \no \\
&=& \mbox{grad }(C^\infty(M)) \oplus \ker \mbox{div}.
\eeqa
Thus
\beq
\ker \mbox{div} = \ker \Delta_H \oplus \mbox{rot }(\Lambda^1(M)),
\eeq
and this is the sought-for refinement of (1.7). The relation (1.12) also
shows that the de~Rham cohomology group $H^2 = \ker \mbox{div/rot }
(\Lambda^1(M))$ is isomorphic to $\ker \Delta_H$. In other words: the
possible failure of the expression (1.8) to furnish the general
solution to Equ. (1.6) is measured by the second Betti number of $M$,
in particular is a topological invariant of $M$. At the same time, 
using the formal self-adjointness of rot and the fact that $\Delta_H$
and rot commute, it follows from the second line of (1.11) that
\beq
\ker \mbox{rot} = \mbox{grad }(C^\infty(M)) \oplus \ker \Delta_H.
\eeq
Thus $H^1 = \ker \mbox{ rot/grad }(C^\infty(M))$ is also isomorphic
to $\ker \Delta_H$, which is an expression of Poincar\'e duality in
the situation at hand. Note that
\beq
\ker \Delta_H = \ker \mbox{div} \cap \ker \mbox{rot}.
\eeq

We now ask whether a similar scenario exists for Equ. (1.1). This,
indeed, turns out to be the case provided $(M,g)$ is (locally)
conformally flat. The associated elliptic complex has been found by
Gasqui and Goldschmidt [10] in a study of infinitesimal deformations
of conformally flat structures for general manifold dimension 
$n \geq 3$. Their work starts from the left end of the complex, i.e.
the conformal Killing equation $LW = 0$. (Their method is to apply the
Spencer--Kodaira--Quillen--Goldschmidt (see e.g. [17]) theory of
overdetermined systems to $LW = 0$.) In a similar vein Calabi [3] and
B\'erard-Bergery et al. [1] had previously considered the integrability
theory of the Killing equation $(M,g)$ when $(M,g)$ is a space of
constant curvature.

\paragraph{Acknowledgements:} At the time of the lecture given at 
the Banach centre I was unaware
of the Gasqui-Goldschmidt  work. 
I am indebted to Professor J.-P. Bourguignon for
pointing out its existence and for helpful discussions. Furthermore
I am grateful to Professor D. Burghelea for teaching me the notion of an
elliptic complex in the early stages of this work and to Professor
J. Lafontaine for important information regarding the premoduli space of
conformally flat structures in the case of space forms. I also thank
Professor L. Andersson for telling me of Ref. [3] and Professor S. Deser
for comments on the manuscript. \\

In the next paragraph of this paper we describe the conformal 
elliptic complex which
plays the same role for TT-tensors as the role played by the 
de~Rham--Hodge complex for Equ. (1.6). Here the operator
rendering explicit solutions (to the TT-condition) is $H$, the
Cotton--York tensor linearized at the conformally metric $g$, viewed
as a map sending trace-free symmetric tensors into themselves. When
$M$ is compact,
the obstruction to tensors in the image of this map to furnish the
general solution of (1.1), by an analogue of Poincar\'e duality,
the same as the obstruction to Killing forms to be the general elements 
of the null space of $H$. This, in turn, has a nice geometric
interpretation: namely it is the premoduli space at $g$ of the space
of conformally flat deformations of $g$. 
For the general, non-compact, case, but when $M$ is assumed to be simply
connected, Gasqui and Goldschmidt [10] have shown that the latter
cohomology, namely $\ker H/L(\Lambda^1(M))$, is zero. In \S~3 of the
present paper we prove our main result. It states that, when $M$ is
simply connected and its second de~Rham cohomology is zero, the
cohomology $\ker \delta/H(S_0^2(M,g))$ vanishes.
In \S~4 we compute this space when $(M,g)$ is a compact space-form.
In the elliptic case the obstruction is found to be zero, i.e. $g$
is infinitesimally rigid as a conformally flat structure. In the flat case,
where $M$ is necessarily a torus, we find the obstruction space is
five-dimensional: this corresponds to flat deformations modulo constant
rescalings of $g$.
In the hyperbolic case the deformation space is given by the space of
tracefree Codazzi tensors on $(M,g)$. This result has already been
obtained by Lafontaine [13] for general dimension $\geq 3$ of $M$

\section{The Conformal Elliptic Complex}
Let now $(M,g)$ be conformally flat, with $M$ not necessarily compact.
Recall that this means that each point of $M$ has a coordinate
neighbourhood $x^a$ in which
\beq
g_{ab} = \omega^2 \delta_{ab}, \qquad \omega > 0,
\eeq
where $\delta_{ab}$ is the flat Euclidean metric. It is well known [15]
that, in dimension 3, this is equivalent to the vanishing of the
Cotton--York tensor $\Ha_{ab}$ defined by
\beq
\Ha_{ab} = 2 \ve_{cd(a} D^c \R^d{}_{b)} = \mbox{rot}_2 \R_{ab}
\eeq
where $\R_{ab}$ is the Ricci curvature of $g_{ab}$. Note the
following properties of $\Ha_{ab}$.
\begin{enumerate}
\item[(i)] $g^{ab} \Ha_{ab} = 0$
\item[(ii)] $D^a \Ha_{ab} = 0$
\item[(iii)] $\Ha_{ab}[\omega^2g] = \omega^{-1} \Ha_{ab}[g]$,
$\omega > 0$.
\end{enumerate}
Geometrically these properties arise as follows. Let $U \subset M$ be a
coordinate neighbourhood and consider the Chern--Simons action
\beq
S[g] = \int_U \ve^{abc} \left( \Gamma_d{}^e{}_a 
\R_{bc}{}^d{}_e - \frac{2}{3} \Gamma_d{}^e{}_a \Gamma_e{}^f{}_b 
\Gamma_f{}^d{}_c \right) \sqrt{g} \; d^3x,
\eeq
where $\Gamma_b{}^a{}_c$ are the Christoffel symbols in the local chart
$x^a$. The functional $S[g]$ has the (non-obvious) properties of being
invariant a) under conformal rescalings of $g$ and b) invariant under
infinitesimal coordinate changes, provided these changes suitably
approach the identity on $\partial U$. (See Chern [4] and Deser et al.
[7].) Note, finally, that the Euler--Lagrange expression of $S[g]$
is nothing but $- 3 \Ha^{ab}$. Then (i,iii) are implied
by a) and (ii) is implied by b).

Consider, next, the linearization of $\Ha_{ab}$ at a conformally flat
metric $g$, i.e. at a metric $g_{ab}$ satisfying $\Ha_{ab}[g] = 0$.
The resultant object, which we call $H(h)$, i.e.
\beq
H_{ab}(h) = \left. \frac{d}{d\lambda}\right|_{\lambda=0} \Ha_{ab}
[g + \lambda h],
\eeq
is a third-order linear partial differential operator acting on
symmetric tensors $h_{ab}$. By virtue of $H$ being the Hessian of $S$
at a critical point, it is formally self-adjoint. Equivalently we can use
the tensor
\beq
\B_{abc} := \ve_{ab}{}^c \; \Ha_{cd} = 2 D_{[a} L_{b]c} ,
\eeq
where $L_{ab} := \R_{ab} - \frac{1}{4} g_{ab} \R$, $\R = g^{ab} \R_{ab}$,
and $B_{abc}$, the linearization of $\B_{abc}$ at $g$.
In the following we shall apply the operators $H_{ab}$ and $B_{abc}$
only to tensors which are trace-free. With this assumption, $B_{abc}$
is explicitly given by
\beq
B_{abc} = 2 (D_{[a} \sigma_{b]c} - C^d_{c[a} L_{b]d}),
\eeq
where 
\beq
\sigma_{ab} = D_{(a} D^c h_{b)c} - \frac{1}{2} \Delta h_{ab} -
\frac{1}{4} g_{ab} D^c D^d h_{cd} + 3 \R_{(a}{}^c h_{b)c} - \frac{3}{4}
g_{ab} h^{cd} \R_{cd} - \frac{3}{4} \R h_{ab}
\eeq
with $\Delta := g^{ab} D_a D_b$ the rough Laplacian and
\beq
C_{ab}^c = \frac{1}{2} g^{cd} (D_a h_{bc} + D_b h_{ac} - D_c h_{ab}).
\eeq

We note the following property of TT-tensors:
\beq
\Delta h \sim - \frac{1}{4} \mbox{ (rot}_2){}^2 h \qquad \mbox{and} \qquad
H h \sim \frac{1}{8} \mbox{ (rot}_2){}^3 h, \qquad \mbox{provided }
\delta h = 0
\eeq
where ``$\sim$'' denotes ``modulo curvature terms''.
Now consider the following sequence
\beq
0 \ra \Lambda^1(M) \st{L}{\ra} S_0^2(M,g) \st{H}{\ra} S_0^2(M,g)
\st{\delta}{\ra} \Lambda^1(M) \ra 0.
\eeq

\paragraph{Proposition} (Gasqui \& Goldschmidt): The sequence (2.10)
is an elliptic complex.

\paragraph{Proof:} Since $g$ satisfies $\Ha_{ab}[g] = 0$, it follows
immediately from property (ii) of $\Ha_{ab}$ that $\delta \circ H = 0$.
But $L = - \delta^*$, and thus
$$
H \circ L = - H \circ \delta^* = - H^* \circ \delta^* =
- (\delta \circ H)^* = 0,
$$
so (2.10) is a complex. Ellipticity at the far left and right of this
complex is equivalent to $\delta$ being an underdetermined elliptic
operator, which we have checked already. Ellipticity at the second and
third place is seen as follows: 
Denote the symbol of any operator $\cO$ by $\bar \cO(k)$. Then, for
example, we want to solve
\beq
\bar H(k)  h = t, \qquad \bar \delta(k) t = 0, \qquad k \neq 0.
\eeq
Using (2.9) we easily see that $\ker \bar H(k) \cap \ker \bar \delta(k) =
\{0\}$. Thus $S_0^2({\bf R}^3,g_x) = \mbox{im } \bar H(k) \oplus \ker
\bar \delta(k) = \ker \bar H(k) \oplus \mbox{im } \bar L(k)$. The last two
relations imply that both (2.11) and
\beq
\bar L(k) W = h, \qquad \bar H(k) h = 0, \qquad k \neq 0
\eeq
have solutions. This ends the proof of the Proposition.

Next observe that all operations in (2.10) are natural under conformal
rescalings of $g$. Thus, when $g' = \omega^2 g$,
\beq
\ba{rclrcl}
L'  W' &=& \omega^2 L W, & \qquad  W' &=& \omega^2 W \\
H'  h' &=& \omega^{-1} H h, & \qquad h' &=& \omega^2 h \\
\delta't' &=& \omega^{-3} \delta t , & \qquad
t' &=& \omega^{-1} t.
\ea 
\eeq
Therefore the cohomology groups associated with (2.10) only depend on $[g]$,
the conformal structure of $g$ which is, of course, locally trivial.
Gasqui and Goldschmidt have shown that, if the above complex is interpreted
in the sense of local formal power series expansions the cohomologies
associated with it, except for the first one, are all trivial.

The first cohomology group of (2.10), namely $\ker L$, is the
space of (globally defined) conformal Killing vectors on $(M,g)$. The second
cohomology group, $\ker H/L(\Lambda^1(M))$, is nothing but the premoduli
space of conformally flat structures around $[g]$. The remaining
cohomologies do not have an immediate geometrical interpretation. In the
compact case, however, we have the following duality.

\paragraph{Theorem:} Let $M$ be compact. Then all cohomologies are
finite-dimensional and
\beq
\ker L \cong \Lambda^1(M)/\delta(S_0^2(M,g)), \qquad
\ker \delta/H(S_0^2(M,g)) \cong \ker H/L(\Lambda^1(M)).
\eeq
\paragraph{Proof:} For the first isomorphism just recall that
$\Lambda^1(M) = \ker L \oplus \delta(S_0^2(M,g))$. The finite dimensionality of
the first summand follows by differential geometry or by noticing
that $\ker L = \ker \delta L$ and $\delta L$ is an elliptic operator.
For the second isomorphism, define the ``generalized Laplacian''
{\bf D} by ${\bf D} = H^2 + (L \delta)^3$. This is a sixth-order, 
elliptic, self-adjoint operator ${\bf D} : S_0^2(M,g) \ra S_0^2(M,g)$. 
Its kernel
is finite-dimensional and given by $\ker {\bf D} = \ker H \cap
\ker \delta$. By an argument analogous to that involving the 
Hodge Laplacian in Sect.~1, one sees that $\ker {\bf D}$ is equal to
both $\ker \delta/H(S_0^2(M,g))$ and $\ker H/L(\Lambda^1(M))$. This ends the
proof of the Theorem.

\section{The Main Theorem}
Gasqui and Goldschmidt [10] have proved the following
\paragraph{Theorem:} Let $M$ be simply connected. Then
$\ker H/L(\Lambda^1(M)) = \{0\}$. \\

Our main result is the following
\paragraph{Theorem:} Let $M$ be simply connected and of vanishing second
de~Rham cohomology. Then $\ker \delta/H(S_0^2(M,g)) = \{0\}$. In other words,
under the above hypotheses, the expression $t = H(h)$ furnishes the
general solution to the equation $\delta t = 0$.

\paragraph{Proof:} The following method of proof is inspired by the proof 
of a Lemma in [16, footnote 13], due to Ashtekar, which states that a tracefree 
Codazzi tensor, i.e. $t_{ab}$ satisfying $\mbox{rot}_2 t = 0$ on a simply
connected space of constant curvature is of the form
$t_{ab} = D_a D_b \alpha - \frac{1}{3} g_{ab} \Delta \alpha$, see also
Ferus [8].
Let $\lambda^a$ be a conformal Killing vector of $(M,g)$. Then, when
$\delta t = 0$, the 2-form defined by $\ve_{ab}{}^d t_{dc} \lambda^c$ is 
curl-free. Thus
\beq
\ve_{ab}{}^d t_{dc} \lambda^c = D_{[a} G_{b]}.
\eeq
Since the space of $\lambda$'s is finite-dimensional there is a way to
linearly assign to each $\lambda$ a covector $G_b$. Pick any such
assignment. Thus we can write $G_a = G_a(\lambda)$. By the conformal
flatness of $g$, there is, locally, the maximum number of $\lambda$'s,
that is to say 10. Furthermore, since $M$ is simply connected (see
e.g. [11]) these $\lambda$'s can be extended to global conformal
Killing vectors. These global conformal Killing vectors can be uniquely
characterized by their conformal ``Killing data'', i.e. the values of
$\lambda^a$, $K^{ab} = D^{[a} \lambda^{b]}$, $D \lambda := D^a \lambda_a$
and $D_b D^a \lambda_a$ at any point of $M$. Thus there are tensor
fields $U_{ab}$, $U_{abc} = U_{a[bc]}$, $V_a$, $V_{ab}$ such that
\beq
G_b(\lambda) = U_{bc} \lambda^c + U_{bcd} K^{cd} + V_b(D\lambda) +
V_{bc} D^c(D \lambda).
\eeq
We now insert this into (3.1) and use the conformal Killing equation
satisfied by $\lambda^a$, i.e.
\beq
D_a \lambda_b = K_{ab} + \frac{1}{3} g_{ab}(D \lambda),
\eeq
and some of its corollaries. Since from now on all calculations are
purely local, it is possible to choose a conformal gauge for $g_{ab}$
so that the curvature is zero. With this in mind, there holds
\beq
D_a K_{bc} = - \frac{2}{3} g_{a[b} D_{c]}(D \lambda)
\eeq
\beq
D_a D_b(D\lambda) = 0.
\eeq
Substituting (3.3,4,5) into (3.1,2) and using that the conformal Killing
data are arbitrary, we obtain
\beqa
\ve_{ab}{}^d \; t_{dc} &=& D_{[a} U_{b]c} \\
0 &=& D_{[a} U_{b]}{}^{cd} + U_{[a}{}^{[c} \delta_{b]}{}^{d]} \\
0 &=& - \frac{1}{3} U_{[ab]} + D_{[a} V_{b]} \\
0 &=& \frac{2}{3} U_{[ab]}{}^c - V_{[a} \delta_{b]}{}^c + 
D_{[a} V_{b]}{}^c .
\eeqa
We solve the equations ``from bottom to top'' except for (3.8) which
turns out to be implied by the remaining relations. Since
$U_{abc} = U_{a[bc]}$, there is the identity
\beq
U_{abc} = U_{[ab]c} + U_{[ca]b} - U_{[bc]a}.
\eeq
Substituting from (3.9) into the right-hand side of (3.10), inserting
into (3.7) and taking a trace of (3.7) we find after some calculation
that
\beqa
U_{bc} &=& 3D_b V_c + \frac{3}{2} g_{bc}(\Delta V_d{}^d - D^dD^e V_{de}) -
\no \\
&& \mbox{} - 3(\Delta V_{(bc)} - D^d D_c V_{(bd)}) + 3D_b D^d V_{(cd)}
- \no \\
&& - 3 D_b D_c V_d{}^d - 3 D_b D^d V_{[cd]}.
\eeqa
Thus the antisymmetric part $V_{[ab]}$ of $V_{ab}$ does not contribute
to $D_{[a} U_{b]c}$. Inserting (3.11) into (3.6), we finally obtain
$(h_{ab} := \frac{3}{8} V_{(ab)} - \frac{1}{8} g_{ab} V_c{}^c)$
\beq
\ve_{ab}{}^d \; t_{dc} = - \frac{1}{8} D_{[a} (\Delta h_{b]c} -
D^d D_{|c|} h_{b]d}) + \mbox{trace-terms}.
\eeq
Thus $V_c{}^c$ drops out of (3.12). Furthermore trace-terms, i.e. 
terms of the form $g_{c[a} \;\cdot_{\;b]}$, do not contribute 
to $t_{ab}$. So taking the dual of (3.12) with respect to the indices
$a$ and $b$, we obtain that $t_{ab}$ has exactly the form of $H_{ab}$,
as given from $H_{ab} = \frac{1}{2} \ve_{(a}{}^{cd} B_{|cd|b)}$ and
Equ.'s (2.6,7) in the local gauge where $\R_{ab} = 0$. Thus we have
proved the Theorem.

For the remaining cohomology in (2.10) we have no results except for
the duality in the compact case.

\section{Compact Space-Forms}
Let $M$ be compact and $g_{ab}$ of constant curvature. In 3 dimensions 
\beq
\R_{abcd} = \frac{\R}{3} g_{c[a} g_{b]d} \Longleftrightarrow
\R_{ab} = \frac{\R}{3} g_{ab}, \qquad \R = \mbox{const.}
\eeq
Clearly $g_{ab}$ is conformally flat. We want to compute the space of
essential infinitesimal conformally flat deformations of $g_{ab}$,
i.e. $\ker H/L(\Lambda^1(M))$. By the Proposition at the end of the previous
section, this amounts to determining $\ker H \cap \ker \delta$.
Using (2.6,7,8) and (4.1) we find after a straightforward calculation
that, when $D^a h_{ab} = 0$,
\beq
B_{abc} = - D_{[a} \Delta h_{b]c} + \frac{\R}{3} D_{[a} h_{b]c}.
\eeq
Furthermore, when $D^a h_{ab} = 0$,
\beq
D_{[a} \Delta h_{b]c} = \Delta D_{[a} h_{b]c}.
\eeq
Thus $\ker H/L(\Lambda^1(M))$ is the same as the space of solutions of the
system
\beq
\left( \Delta - \frac{\R}{3}\right) D_{[a} h_{b]c} = 0, \qquad
D^a h_{ab} = 0.
\eeq
Suppose first that $\R \geq 0$. Then, contracting the first equation in
(4.4) with $D^{[a} h^{b]c}$ and integrating by part, we find that
\beq
D_d D_{[a} h_{b]c} = 0.
\eeq
Contracting (4.5) with $g^{da}$ this implies
\beq
\left( \Delta - \frac{\R}{2}\right) h_{ab} = 0.
\eeq
Upon contraction with $h^{ab}$ and integration this implies
\beq
D_a h_{bc} = 0
\eeq
and
\beq
h_{ab} = 0 \qquad \mbox{when } \R > 0.
\eeq
Thus, in the elliptic case (i.e. $\R > 0$), the conformal structure
defined by $g_{ab}$ is rigid amongst all conformally flat structures
on $M$. When $\R = 0$, $(M,g)$ has to be a flat torus ${\bf T}^3$ 
(see Ch. V, Theorem 4.2 of [12]).
Corresponding to each ${\bf S}^1$-factor of ${\bf T}^3$ there is a
covariantly constant vector. Taking tensor products, symmetrizing and
subtracting out the trace, we obtain a 5-parameter set of tensors
obeying (4.7). Since solutions to (4.7) are uniquely determined by
their value at some point, these are all solutions to Equ. (4.7).
Thus these deformations stay within metrics on ${\bf T}^3$ conformal to
a standard flat one.

Finally we consider the hyperbolic case, $\R < 0$. We use the identity,
valid when $D^a t_{ab} = 0$,
\beq
\Delta t_{ab} = - \frac{1}{4} (\mbox{rot}_2)^2 t_{ab} + \frac{\R}{2}
t_{ab}.
\eeq
Thus (4.4) implies
\beq
\left[ (\mbox{rot}_2)^2 - \frac{2\R}{3}\right] \mbox{ rot}_2 t_{ab}
= 0.
\eeq
Contracting (4.10) with rot$_2 t^{ab}$ and integrating, noting that 
rot$_2$ is self-adjoint and using $\R > 0$, yields
\beq
\mbox{(rot}_2)^2 \; t_{ab} = 0,
\eeq
which, upon contracting with $t^{ab}$, results in
\beq
\mbox{rot}_2 \; t_{ab} = 0 \Longleftrightarrow D_{[a} t_{b]c} = 0.
\eeq
Thus, in the hyperbolic case, the infinitesimal deformation space
$\ker H/L(\Lambda^1(M))$ is isomorphic to the space of traceless 
Codazzi tensors. This, as Lafontaine [13] has shown, {\bf can} be
non-trivial for certain space forms. By the Mostow rigidity
theorem [14], the deformed conformally flat structures can not
again be space forms. In fact such deformations (even finite ones)
{\bf have} been constructed (see refs. in [13]), 
using methods completely beyond the ones of this paper.

\end{document}